\documentclass{webofc}
\usepackage[varg]{txfonts}   
\usepackage[mathlines]{lineno}
\begin{document}
\title{Light and strange hadron production and anisotropic flow measurement in Au+Au collisions at ${\sqrt{s_{\rm NN}} = \rm{3\,GeV}}$ from STAR }
%
%

\author{\firstname{Guannan} \lastname{Xie (for the STAR Collaboration)}\inst{1}\fnsep\thanks{\email{GuannanXie@lbl.gov}} 
}

\institute{Lawrence Berkeley National Laboratory}

\abstract{%
  In this proceeding, we present on our first measurements of identified particle ($\pi$, $K$, $p$, $K_{s}^{0}$, $\Lambda$, $\phi$, $\Xi^{-}$) production and anisotropic flow ($v_{1}$, $v_{2}$) in Au+Au collisions at ${\sqrt{s_{\rm NN}} = \rm{3\,GeV}}$. Various models including thermal and transport model calculations are compared to data, these results imply that the matter produced in the 3 GeV Au+Au collisions is considerably different from that at higher energies.

}
\maketitle
\section{Introduction}
\label{intro}

  Relativistic heavy ion physics is aiming at the detailed investigation of phase structures of strongly interacting matter under extreme conditions. Searching for the onset of Quark-Gluon Plasma (QGP), studying the properties of the produced QCD matter, and locating the possible QCD phase boundary are the focus of the RHIC Beam Energy Scan (BES) Program~\cite{HotQCD_HeavyIon}.

Particle production and collective flow have been used to investigate the properties of the QCD matter produced in heavy-ion collisions. The RHIC BES program covers a wide range of energies, including the transition from a partonic dominated area to hadronic dominated area. Of particular interest is the high baryon density region which is accessible through production and flow measurements of particles including light and strange hadrons in the STAR fixed-target program. In these proceedings, invariant yields and rapidity density distributions of $\pi$, $K$, $\phi$ mesons and $\Xi^{-}$ hyperons as well as the directed/elliptic flow of $\pi$, $K$, $p$, $K_{s}^{0}$, $\Lambda$, $\phi$ are presented. 

\section{Experiment}
\label{sec-1}

The dataset used in this analysis consists of Au+Au collisions at ${\sqrt{s_{\rm NN}} = \rm{3\,GeV}}$ collected by the STAR experiment under the fixed target (FXT) configuration in the year of 2018. The single beam was provided by RHIC with total energy equal to 3.85 GeV/nucleon. The thickness of the gold target is about 0.25 mm, corresponding to a 1\% interaction probability to minimize the pileup and energy loss effect in target. The target is located at 200\,cm to the west of the center of the STAR detector, it is installed inside the vacuum pipe, 2\,cm below the center of the beam axis. The main detectors used for this analysis are the Time Projection Chamber (TPC), the Time of Flight (TOF) detector, and the Beam-Beam Counter (BBC)~\cite{STAR}. The trigger is provided by the signal in the east BBC detector and at least five hits in the TOF detector. Tracking and particle identification (PID) are done using the energy loss (dE/dx) information from TPC and time of flight (1/$\beta$) information from TOF. In total, approximately 2.6$\times 10^{8}$ minimum bias triggered events are used in this analysis. 
Reconstruction of short lived particles, $K_S^{0} \rightarrow \pi^{+}+\pi^{-}$, $\Lambda \rightarrow p+\pi^{-}$ and $\Xi^{-} \rightarrow \Lambda+\pi^{-}$ is performed using the KF Particle Finder package based on the Kalman Filter method~\cite{STAR}. Those combinatorial backgrounds are obtained by rotating daughter tracks. The $\phi$ mesons are reconstructed through the hadronic decay channel, $\phi \rightarrow K^{+}+K^{-}$, where combinatorial background is estimated using the mixed event technique. After corrected for the acceptance and efficiency, the 4$\pi$ acceptance yield (dN) and anisotropic flow ($v_1, v_2$) are obtained.



\section{Light and Strange Hadron Yields}
\label{sec-2}

\begin{figure}[h]
\centering
\includegraphics[width=12cm,clip]{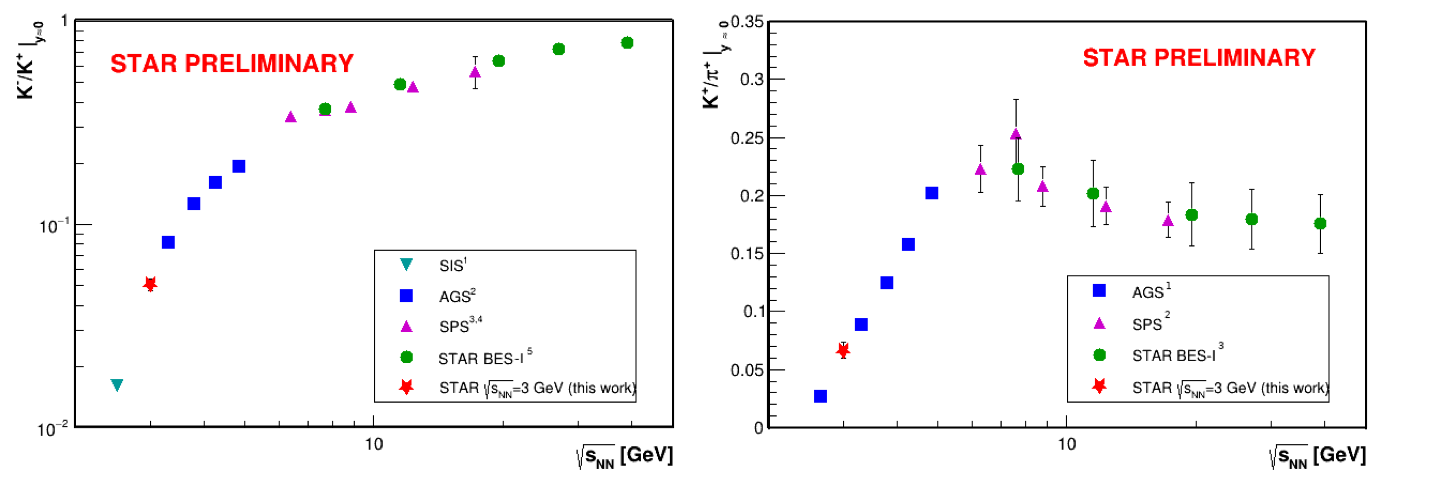}
  \caption{$K^-/K^+$ (left) and $K^+/\pi^+$ (right) ratio as a function of $\sqrt{s_{\rm NN}}$. The red star shows the measurements at 3 GeV with statistical and systematic uncertainties added together, while other markers are used for data from other energies~\cite{pi2K_experiment}.} 
\label{fig-1}       
\end{figure}

Figure~\ref{fig-1} left shows the $K^-/K^+$ ratio along with collision energy ${\sqrt{s_{\rm NN}}}$. The relative yield of $K^-$ is much smaller compared to $K^+$ at 3 GeV which demonstrates the importance of $K^+$ production in association with the $\Lambda$ ($N+N \rightarrow N+\Lambda+K$). Right plot shows the $K^+/\pi^+$ ratio along with ${\sqrt{s_{\rm NN}}}$. The horn structure around 7 GeV was proposed as a possible signal of onset of deconfinement in early days, but later on the data are described well by the statistics models. The measured $K^-/K^+$ and $K^+/\pi^+$ ratios at 3 GeV follow with the world trend~\cite{pi2K_experiment}.

\begin{figure*}
\centering
\includegraphics[width=5.cm,clip]{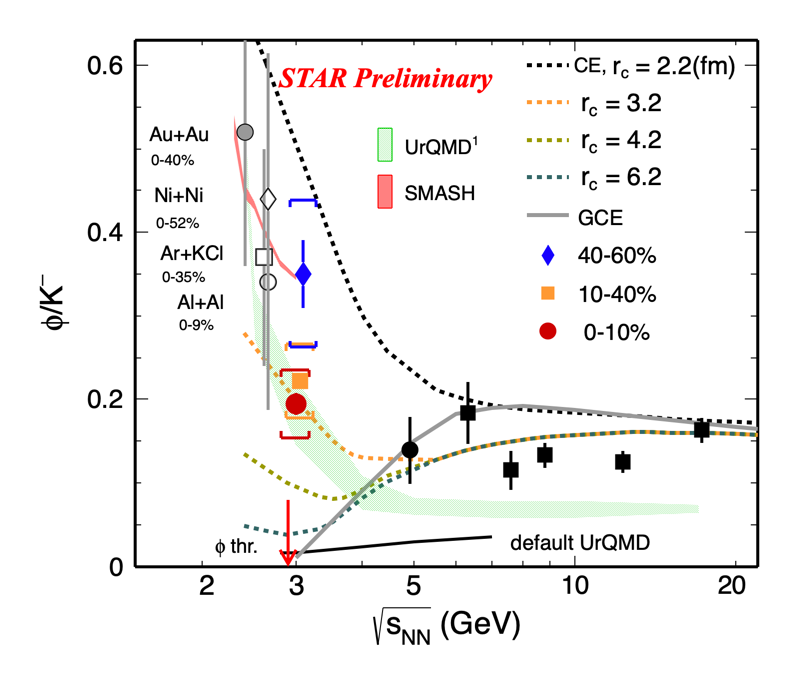}
\includegraphics[width=5.cm,clip]{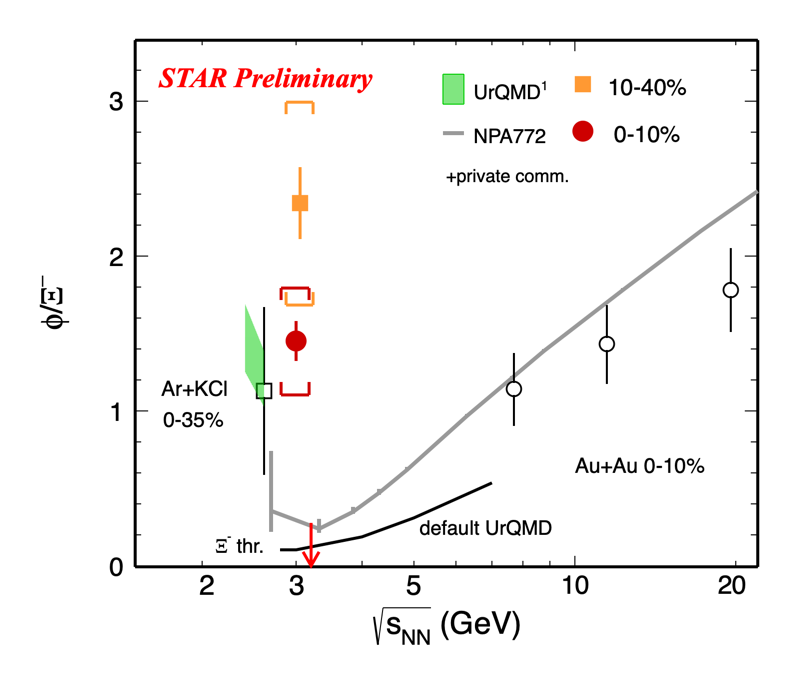}
  \caption{$\phi/K^-$ (left) and $\phi/\Xi^-$ (right) ratio as a function of $\sqrt{s_{\rm NN}}$. The solid color markers show the measurements at 3 GeV from different centrality bins with vertical lines for statistical and bracket symbols for systematic errors, while other markers are used for data from various other energies and/or collision systems~\cite{phi2K_experiment}.} 
\label{fig-2}       
\end{figure*}

Statistical thermal models have often been used to characterize the thermal properties of the produced media. In low energies, the strangeness production is rare, therefore it has been argued that strangeness number needs to be conserved locally on an event-by-event basis described by the Canonical Ensemble (CE), which leads to a reduction in the yields of hadrons with non-zero strangeness number. Figure~\ref{fig-2} shows the $\phi/K^-$ (left) and $\phi/\Xi^-$ (right) ratios from our measurements in different centralities as a function of $\sqrt{s_{\rm NN}}$. The measured $\phi/K^-$ and $\phi/\Xi^-$ ratios at 3\,GeV are slightly higher than the values at high energies for $\sqrt{s_{\rm NN}}\geqslant$ 7\,GeV. There is a hint for both the measured $\phi/K^-$ and $\phi/\Xi^-$ ratios in mid-central collisions are larger than that in central collisions, and this need further statistics to systematically study the centrality dependence in detail. The Grand Canonical Ensemble (GCE) underestimates our data with $\sim$5$\sigma$ effect for $\phi/K^-$ and $\sim$4$\sigma$ effect for $\phi/\Xi^-$, while the CE calculation with strangeness correlation length ($r_c$) $\sim$3.2 fm can reasonably describe our measurements. The precise determination of the thermal parameters (including $T_{ch}$, $\mu_{B}$ and $r_c$) needs a global thermal model fit with all the particle yields at 3 GeV. The modified transport models (UrQMD and SMASH) calculations by including high mass resonances decay to $\phi$ and $\Xi^-$ can also reasonably describe the data at low energies~\cite{phi2K_experiment}. In the Au+Au collisions at 3\,GeV, the observed strangeness production mechanism may be different from that at high energy, and this may indicate a change of EoS at this low energy.

\section{Light and Strange Hadron Anisotropic Flow}
\label{sec-3}

\begin{figure}[h]
\centering
\includegraphics[width=9.8cm,clip]{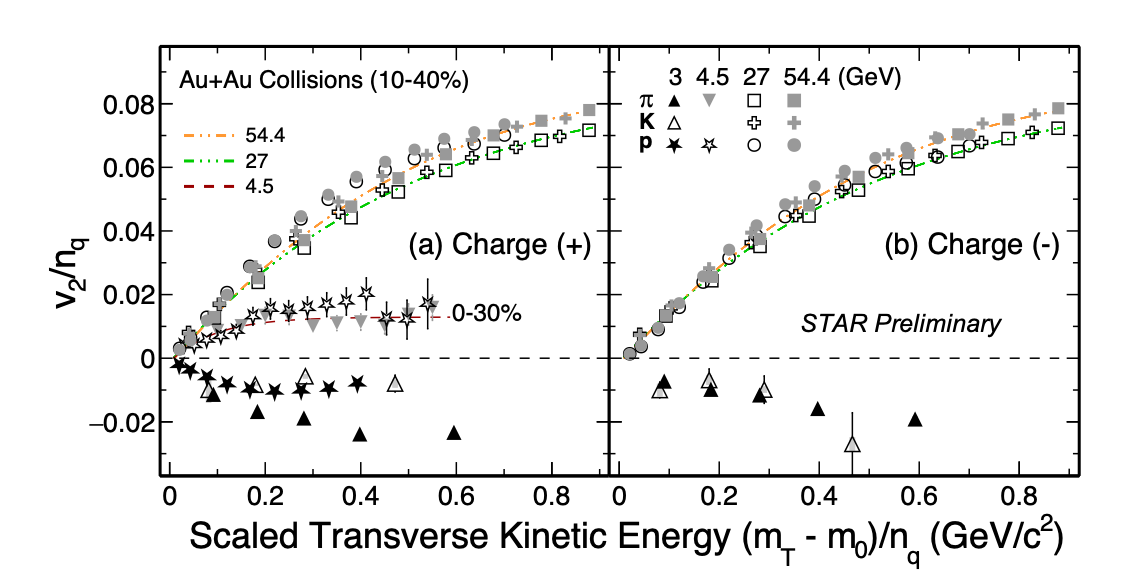}
  \caption{NCQ scaled $v_2$ as a function of $(m_{T}-m_{0})/n_{q}$ for pions, kaons and protons from Au+Au collisions in 10-40\% centrality for positive (left) and negative (right) charged particles.}
\label{fig-3}       
\end{figure}

\begin{figure}[h]
\centering
\includegraphics[width=9.5cm,clip]{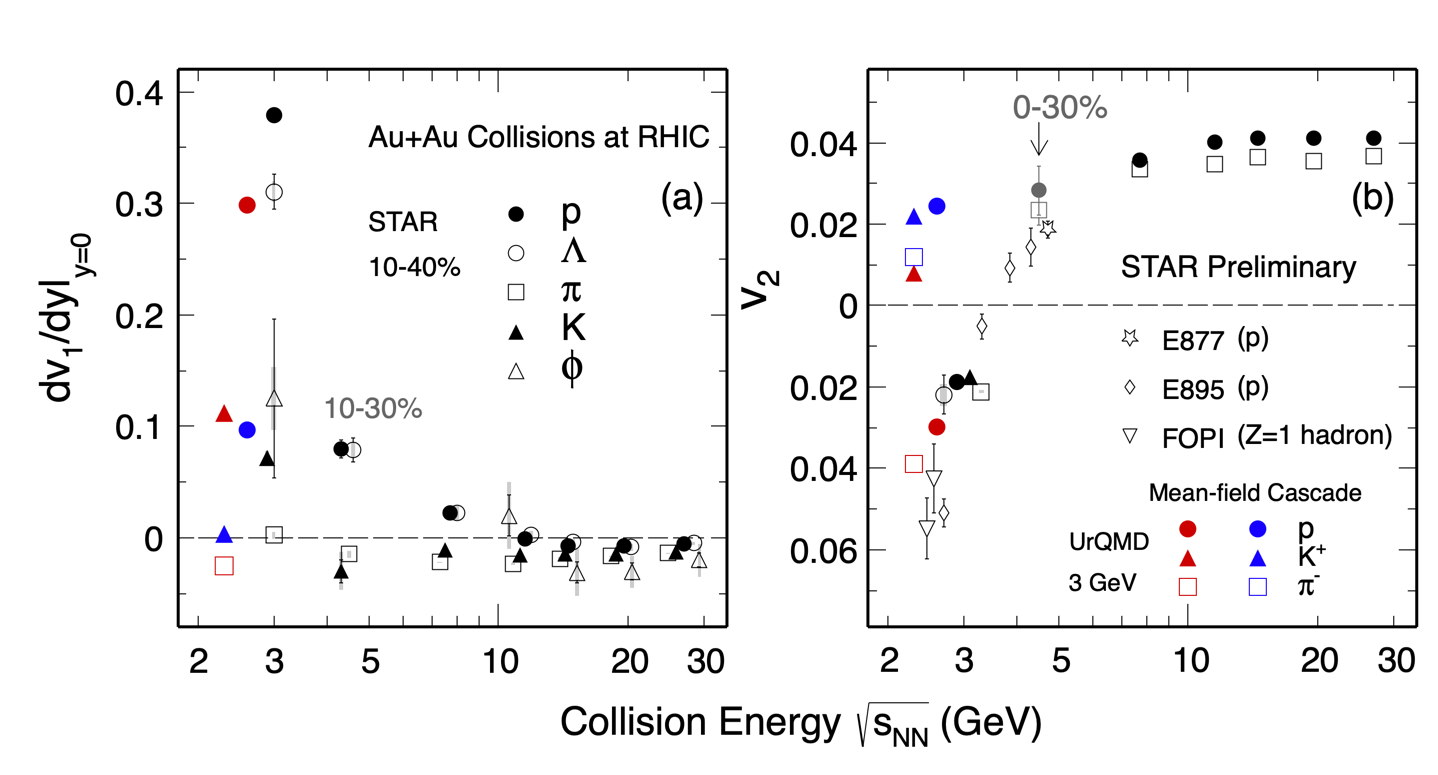}
  \caption{$v_1$ slope $dv_{1}/dy|_{y=0}$ (left) and $v_2$ (right) along with ${\sqrt{s_{\rm NN}}}$ in heavy-ion collisions~\cite{v1v2_experiment}.}
\label{fig-4}       
\end{figure}

Directed flow $v_{1}$ and elliptic flow $v_2$ are sensitive to the stiffness of the nuclear Equation of State (EoS) at high baryon density region~\cite{v1v2_experiment}. The number of constituent quark (NCQ) scaling of $v_2$ is observed for particle or anti-particle for collision energy $\ge$ 7.7\,GeV and it is often argued as an evidence for the formation of a QGP phase with partonic degrees of freedom. For 3\,GeV collisions, the elliptic flow of all charged hadrons measured at midrapidity are negative and the NCQ scaling is absent as shown in Fig.~\ref{fig-3}, especially for positive charged particles. The slope of the midrapidity directed flow of all measured hadrons are found to be positive as shown in Fig.~\ref{fig-4}~\cite{v1v2_experiment}. This is the first time observed of positive $v_1$ slop for kaons and $\phi$ mesons. In addition, the UrQMD model calculations with the baryonic mean-field potential qualitatively reproduce our observations. Due to the formation of the QGP, at ${\sqrt{s_{\rm NN}}}$ larger than 10\,GeV, we observed the opposite behaviors, namely all hadrons' $v_2$ are positive while all slopes of $v_1$ are negative. In the Au+Au collisions at 3\,GeV, the observed opposite behavior implies the vanishing of partonic collectivity and a new EoS dominated by the baryonic interactions in these region.

\section{Summary}
\label{sum}

In Au+Au collisions at 3 GeV, our result suggests that the strangeness production mechanism may be different compared to that in high energies. The collective flow behaviors show an opposite trend to that in high energies. These results imply that the matter produced in the 3 GeV Au+Au collisions is considerably different from those at higher energies.

%
%
%

\end{document}